\documentclass{nature}

\usepackage{graphicx}
\usepackage{amsmath}
\usepackage{booktabs}

\let\NATgraphics\includegraphics
\AtBeginDocument{\let\includegraphics\NATgraphics}

\title{Vertical breathing in the Milky Way's bar region follows the bar}

\author{Zhen Wan$^{1,2}$ \& Lulu Fan$^{1,2,3,4}$}

\begin{document}

\maketitle

\begin{affiliations}
 \item Department of Astronomy, University of Science and Technology of China, Hefei 230026, China
 \item School of Astronomy and Space Science, University of Science and Technology of China, Hefei 230026, China
 \item Deep Space Exploration Laboratory, Hefei 230088, China
 \item College of Physics, Guizhou University, Guiyang 550025, China
\end{affiliations}

\begin{abstract}
Bars and spiral arms can make a stellar disc breathe, with stars on both sides of the plane moving away from it or towards it together. Breathing has been measured from the outer disc to the end of the bar, but not across the bar region. Here we use about five million Gaia DR3 stars with radial velocities to measure the part of the latitudinal proper motion that changes sign with latitude, and compare it with an N-body bar observed with the same selection and noise. The bar region breathes: the near side is compressing ($-4.3\sigma$ and $-4.4\sigma$ relative to no vertical motion, in two distance ranges), the centre is expanding ($6.0\sigma$) and control longitudes show no signal. A bar with a pattern speed of 38.5\,km\,s$^{-1}$\,kpc$^{-1}$ reproduces this sequence of signs (correlation 0.84; data amplitude 0.83 times the model); the same disc before the bar formed shows no such pattern (correlation $-0.16$). On the far side, a weaker expansion ($3.1\sigma$) begins at $l\approx-9^\circ$, compared with $l\approx-17^\circ$ in the model, although this far-side difference is not robust under our predefined criterion. Vertical motions in the bar region thus offer a direct test of bar models.
\end{abstract}

When a stellar disc breathes, stars on the two sides of the Galactic plane move away from the plane together, which thickens the disc, or towards it together, which thins it. Near the Sun, the mean vertical velocity of disc stars shows large-amplitude structure, up to 17\,km\,s$^{-1}$, with signs of alternating compression and expansion\cite{Williams2013}. With Gaia DR2 and DR3, the Gaia Collaboration separated this north--south antisymmetric motion from bending, in which both sides move in the same direction, and mapped the two across the disc\cite{Gaia2018,Gaia2023disc}. The bulk vertical motions of the outer disc are consistent with a combination of breathing and bending\cite{Carrillo2019}, and compressing breathing motions have been found along the Local Arm and in the inner Galaxy near the end of the bar\cite{Asano2024}. Stars near the Sun on orbits from the inner disc also show two-armed spirals in the plane of vertical position and velocity, a signature of breathing in phase space\cite{Hunt2022,Li2023b}.

Bars and spiral arms can drive breathing. A single bar or spiral pattern is enough to generate bulk vertical motions in a disc\cite{Monari2015}. When a bar and spiral arms act together, the breathing motions are a non-linear superposition of their separate effects and considerably exceed the sum of the two\cite{Monari2016}. In test-particle models with a steadily rotating bar and two-armed spirals, these motions reach a few km\,s$^{-1}$\cite{Li2023b}. In simulations, breathing waves rotate with the spirals that drive them, and their pattern speeds can also match that of the bar\cite{Khachaturyants2022}. When a bar buckles, vertical oscillations arise and wind up into spirals in phase space\cite{Khoperskov2019}. The boxy/peanut shape of the Milky Way's bulge indicates that its bar probably buckled\cite{Lokas2019}.

The bar region is thus a natural place to look for breathing. Existing observations constrain the proper motions of red clump stars in the bulge\cite{Vasquez2013}, the three-dimensional kinematics of the bar and bulge\cite{Clarke2019,Sanders2019,Zhang2024} and the chemo-kinematics of the inner disc\cite{Queiroz2021}. Face-on maps of the mean vertical velocity show no systematic pattern on either side of the bar\cite{Gaia2023disc,Zhang2024}, but breathing moves stars on the two sides of the plane in opposite vertical directions and cancels when the two sides are averaged. The mean latitudinal proper motions of bulge stars point towards the plane on the near part of the bar and away from it on the far part, a pattern that signals the rotation of the bar and streaming motions within it\cite{Clarke2019}. To our knowledge, no measurement across the bar region has yet isolated the mean vertical motion of stars on each side of the plane from these motions within the plane. Here we measure breathing in the bar region from Gaia DR3 stars with radial velocities. From the latitudinal proper motion of each star we subtract the value it would have without vertical motion, computed from its radial velocity and parallax, and we use the part of the residual that changes sign with Galactic latitude. We find that the near side of the bar region is compressing and the centre is expanding (Fig.~1). An N-body bar with a pattern speed within the measured range for the Milky Way, observed with the same selection and noise, produces the same sequence of signs with a comparable amplitude, and the same disc before the bar formed shows no such pattern. On the far side, the data turn to expansion closer to the centre than the model does, although this difference is not robust under our predefined criterion.

\section*{Results}

\subsubsection*{Near-side compression and central expansion}

Breathing in the bar region appears in the latitudinal proper motion, $\mu_b$, as a component that changes sign with Galactic latitude, $b$ (Extended Data Fig.~1). For each star we use its own radial velocity and parallax to compute the $\mu_b$ it would have without vertical motion, and subtract this from the observed value. We call the part of the residual that changes sign with latitude the breathing proper motion, $S$. It is positive when the disc expands, because the residual then has the same sign as $b$, and negative when the disc compresses. Breathing near the Sun is measured in km\,s$^{-1}$, because those stars are close and their distances are accurate. In the bar region the distance errors are large. Converting $\mu_b$ into a velocity there multiplies the parallax error by a term that changes sign with latitude, and this term cannot be separated from a true vertical motion. We therefore keep $S$ in mas\,yr$^{-1}$ in Fig.~1 and throughout, and observe an N-body bar with the same sky area, selection and noise before measuring the same quantity. The same model observed with the vertical velocity of every particle removed shows what the selection, the noise, the statistic and any projection of motions within the plane that remains after the subtraction produce on their own; we quote significances relative to this model without vertical motion. The $5\sigma$ rejection of outlying stars removes very few stars and does not change the result (Extended Data Fig.~2).

The near side is compressing, the centre is expanding and the control longitudes show no signal (Fig.~1 and Table~1). In Fig.~1 the Gaia values are negative on the near side and positive at the centre, with a shallow negative stretch just beyond the centre and a return to positive values at more negative longitudes. Table~1 gives the corresponding regional values; each significance quoted here is a $z$ value from Table~1, the difference between two columns divided by its uncertainty, with the sign kept. On the near side ($l=4^\circ$ to $16^\circ$), $S=-0.057$\,mas\,yr$^{-1}$ at parallax distances of 6--15\,kpc, $-4.4\sigma$ relative to no vertical motion, and the significance at 3--6\,kpc is $-4.3\sigma$. The centre ($l=-4^\circ$ to $4^\circ$) is expanding at 6--15\,kpc, with $S=+0.073$\,mas\,yr$^{-1}$ at $6.0\sigma$. The control longitudes ($|l|\geq24^\circ$) give $+1.2\sigma$ and $-0.3\sigma$, consistent with zero. The far side ($l=-20^\circ$ to $-8^\circ$) gives $S=+0.050$\,mas\,yr$^{-1}$ at 6--15\,kpc ($3.1\sigma$), which our predefined criterion classes as marginal, although all six of its $2^\circ$ longitude bins are positive. The model without vertical motion gives an almost flat curve (orange dashed line in Fig.~1), so the selection and the statistic do not create the pattern; it comes from the vertical motion of the stars. Split by parallax distance, the regions show no such sequence of signs within 3\,kpc (Extended Data Fig.~3). Each of four $2^\circ$ latitude bands between $|b|=2^\circ$ and $10^\circ$ shows the same sequence of signs along longitude (Extended Data Fig.~4). Changing the parallax zero-point, the magnitude or colour range, or the number of Gaia scans leaves the signs unchanged within their uncertainties (Extended Data Fig.~5).

\subsubsection*{The pattern comes with the bar}

An N-body bar produces the same sequence of signs with a comparable amplitude, and the same disc shows no such pattern before the bar forms. We built the N-body model from the Milky Way mass model of Tepper-Garc\'ia et al.\cite{TepperGarcia2021} (Methods). Measurements of the pattern speed of the Milky Way's bar\cite{Bovy2019,ClarkeGerhard2022} range from 33 to 41\,km\,s$^{-1}$\,kpc$^{-1}$, so the main comparison uses the model at 2.9\,Gyr, when its pattern speed is 38.5\,km\,s$^{-1}$\,kpc$^{-1}$. In Fig.~1 this model follows the same sequence of signs as Gaia along longitude. Over the 24 longitude bins at $|l|<24^\circ$, the correlation between data and model is 0.84. Allowing for the sampling uncertainty of the model, the data amplitude is 0.83 times that of the model (68\% interval 0.76--0.92). The four regional values on the near side and at the centre differ from the model by at most $0.99\sigma$ (Table~1). Before the bar forms, the same disc at 0.5\,Gyr, processed in exactly the same way, gives a curve whose correlation with the data is $-0.16$ (purple dotted line in Fig.~1). At five epochs of the same simulation, with pattern speeds of 41.9, 38.5, 35.9, 30.6 and 27.3\,km\,s$^{-1}$\,kpc$^{-1}$, the correlations are 0.79, 0.84, 0.67, 0.37 and 0.27 (Extended Data Fig.~6). Pattern speed, bar length and bar strength evolve together in one simulation, so these values show only that the slower, older bars match the data less well; they cannot isolate the effect of the pattern speed. Two differences remain. In the central 3--6\,kpc group the data give $-0.0135$ and the model $+0.0175$\,mas\,yr$^{-1}$; both are close to zero ($z=-0.67$), but their signs differ, so the predefined criterion for agreement, which also requires the same sign, is formally not met (Methods). The next section explains this difference. In the far-side 6--15\,kpc group the regional value of the model lies $2.44\sigma$ below the data, because the data and the model turn to expansion at different longitudes.

\subsubsection*{The signal begins inside the bar region}

The signal begins where the sightlines enter the bar region (Fig.~2). Fig.~2a shows the mean vertical motion of the 2.9\,Gyr model seen from above. At 6--15\,kpc, near-side sightlines cross a compressing zone beside the near end of the bar, which makes $S$ negative, and central sightlines cross two expanding lobes, which make it positive. Fig.~2b resolves $S$ in parallax distance. Within 3\,kpc the sequence of signs is absent. On the near side, the data are negative from 4\,kpc and the model from 3\,kpc; at the centre, the data are positive from 5\,kpc and the model from about 4\,kpc. In both regions the signal therefore appears once the sightlines reach the bar region. The central sign change also explains the 3--6\,kpc difference noted above. It falls inside this group, about 1\,kpc farther for the data than for the model, so both group averages lie near zero with opposite signs. In the model, far-side sightlines at 6--15\,kpc cross the compressing zone beside the far end of the bar; the data and the model differ there, as described next.

\subsubsection*{The far side turns to expansion closer to the centre}

On the far side, the data turn from compression to expansion at $l\approx-9^\circ$, closer to the centre than in the model (Fig.~1). At 6--15\,kpc the data average $+0.047$\,mas\,yr$^{-1}$ between $l=-15^\circ$ and $-9^\circ$ and $+0.065$\,mas\,yr$^{-1}$ between $-23^\circ$ and $-17^\circ$. The 2.9\,Gyr model also expands between $-23^\circ$ and $-17^\circ$, with a similar amplitude of $+0.044$\,mas\,yr$^{-1}$. Its compression, however, extends to $l\approx-15^\circ$ and turns to expansion only at $-17^\circ$, so that between $-15^\circ$ and $-9^\circ$ the model averages $-0.026$\,mas\,yr$^{-1}$. The predefined far-side region ($l=-20^\circ$ to $-8^\circ$) mostly covers this stretch, so the regional value of the model lies $2.44\sigma$ below the data (Table~1). We defined the turnover longitude after seeing the data and use it only as a description. At nine epochs between 2.3 and 3.9\,Gyr and at bar angles from $15^\circ$ to $45^\circ$, the model turns to expansion between $l\approx-29^\circ$ and $-13^\circ$, or not at all within $-30^\circ$ (at 3.1 and 3.9\,Gyr), so no epoch or bar angle turns over as close to the centre as the data. For every bar angle from $15^\circ$ to $45^\circ$, the far-side data at 6--15\,kpc lie above the model (Extended Data Fig.~7). A public snapshot of the original, higher-resolution simulation of the same model\cite{TepperGarcia2021}, at 4.28\,Gyr, has the same signs as the data on the near side and at the centre, with the data at 0.27 times its amplitude, and it is also negative on the far side (Extended Data Fig.~8). In this family of models, therefore, the compressing zone beside the far end of the bar extends to more negative longitudes than in the data. The regional difference nevertheless reaches $|z|\geq3$ in only two of seven comparisons across epochs and bar angles, so it is not robust under our predefined criterion (Methods). Relative to the model, the far-side residual is best described by a single offset that does not change with distance, so part of it could also be a sky-fixed proper-motion error (Methods). The far-side signal itself, relative to the model without vertical motion, depends on distance, as do the near-side and central signals.

\section*{Discussion}

In the model, breathing in the bar region is a lasting product of the bar. The near side compresses and the centre expands at 6--15\,kpc at every epoch we examined between 2.3 and 5.9\,Gyr, long after the bar thickened into a boxy/peanut shape at about 2\,Gyr, and the higher-resolution public snapshot shows the same signs at 4.28\,Gyr. Breathing driven by a bar is expected to rotate with the bar\cite{Monari2015,Khachaturyants2022}. Outside the bar, linear theory\cite{Monari2015} gives bar-driven vertical motions below 1\,km\,s$^{-1}$; inside the bar region of our model, they reach about 10\,km\,s$^{-1}$ (Fig.~2a). The Milky Way follows the sequence of signs of the model, and in both the signal begins where the sightlines enter the bar region (Fig.~2b). Asano et al.\cite{Asano2024} found a compressing zone in the inner Galaxy near the end of the bar and suggested that the bar may drive it; in our model, the near-side compression comes from a zone beside the near end of the bar. The region where breathing follows the bar coincides with the boxy/peanut bulge, which probably formed when the bar buckled\cite{Lokas2019}. What we measure is the present-day vertical motion; connecting it with past buckling requires following the evolution of the bar in models.

These measurements give bar models three quantities that can be compared directly with Gaia: the sequence of signs of the breathing, its amplitude and the far-side turnover longitude. Our model reproduces the sequence of signs, and the data amplitude is 0.83 times that of the model. The public snapshot, at a higher resolution and 1.4\,Gyr later in its evolution, has the same signs with about three times our amplitude, so the amplitude can distinguish models that share the sequence of signs. In the data, the far-side turnover lies closer to the centre than at any epoch or bar angle of this family of models.

Within their uncertainties, the signs do not change under any analysis choice or subsample we tested, and the regional uncertainties include the spatially correlated errors of Gaia proper motions. Three limitations remain, and coming data and models can address each of them. Gaia stars with radial velocities are bright, and extinction keeps fainter stars out of the sample, which is smallest on the far side; SDSS-V\cite{Kollmeier2017} and MOONS\cite{Cirasuolo2020} in the near-infrared, and 4MOST\cite{deJong2019}, will add spectroscopy of fainter stars in the inner Galaxy. The spatially correlated errors of Gaia proper motions exceed the statistical uncertainties of the near-side and central values relative to no vertical motion, and part of the far-side difference could be a sky-fixed proper-motion error; Gaia DR4, based on 66 months of observations instead of 34, is expected to reduce these errors and to provide radial velocities for more stars. Our model uses a single set of initial conditions\cite{TepperGarcia2021}, and with one public snapshot the effects of resolution and epoch on the amplitude cannot be separated; testing them, and other initial conditions, requires further simulations.

\begin{methods}

\subsection{Gaia sample.}
We use Gaia DR3\cite{Gaia2016,GaiaDR3} stars with radial velocities\cite{Katz2023} at Galactic longitudes $|l|<51^\circ$ and latitudes $2^\circ\leq|b|<10^\circ$. Every star has $G<19$, parallax and proper-motion uncertainties below 1\,mas and 1\,mas\,yr$^{-1}$, a renormalized unit weight error (RUWE) below 1.4, a radial-velocity uncertainty below 10\,km\,s$^{-1}$ from at least five transits, and a positive parallax. The catalogue contains 5,627,950 such stars. We correct the radial velocities of stars with $G_{\rm RVS}\geq11$ and template temperatures below 8,500\,K for the magnitude-dependent zero-point\cite{Katz2023} and add $\varpi_0=0.025$\,mas to all parallaxes to correct for the Gaia parallax zero-point\cite{Lindegren2021}. As tests, we repeat the measurement without the radial-velocity correction and with $\varpi_0=0.015$ and 0.035\,mas (Extended Data Fig.~5); the uncertainties also include the effect of a 0.01\,mas error in $\varpi_0$ (Uncertainties). The uncertainty of $\mu_b$ is $\sigma_{\mu_b}=[(\sigma_{\mu_{\alpha*}}^2+\sigma_{\mu_\delta}^2)/2]^{1/2}$, without the correlation term.

\subsection{Breathing statistic.}
For a star at latitude $b$, the velocity along the latitude direction is
\begin{equation}
v_b=-v_r\tan b+\frac{v_z}{\cos b},
\end{equation}
where $v_r$ is the heliocentric radial velocity and $v_z$ the vertical velocity relative to the Sun. Without vertical motion of the star, $v_z=-W_\odot$ and the expected proper motion is
\begin{equation}
\mu_{b,0}=\left(-v_r\tan b-\frac{W_\odot}{\cos b}\right)\frac{\varpi+\varpi_0}{K},
\end{equation}
where $\varpi$ is the parallax in mas and $K=4.74047$\,km\,s$^{-1}$\,kpc$^{-1}$\,(mas\,yr$^{-1}$)$^{-1}$. The solar motion is $(U,V,W)_\odot=(11.1,12.24,7.78)$\,km\,s$^{-1}$, with $U_\odot$ and $V_\odot$ from Sch\"onrich et al.\cite{Schonrich2010} and $W_\odot$ the default value in Astropy\cite{Astropy2013,Astropy2018,Astropy2022}, which follows from the proper motion and distance of Sgr\,A*\cite{DrimmelPoggio2018,Gravity2018,ReidBrunthaler2004}. The statistic uses neither the Galactocentric distance of the Sun nor its height above the plane. In each cell, we fit the residual $\Delta\mu_b=\mu_b-\mu_{b,0}$ as
\begin{equation}
\Delta\mu_b=\beta_s\sin b+\beta_c\cos b
\end{equation}
by unweighted least squares. We do not weight by the formal proper-motion uncertainties, because the scatter of the residuals comes mainly from parallax and radial-velocity errors propagated through $\mu_{b,0}$. Stars that deviate from the fit by more than $5\times1.4826$ times the median absolute residual are rejected, and the fit is repeated up to ten times; in the two cells of Extended Data Fig.~2 this removes one star each. The breathing proper motion is
\begin{equation}
S=\beta_s\frac{\langle\sin^2 b\rangle}{\langle|\sin b|\rangle},
\end{equation}
and its uncertainty follows from 1.4826 times the median absolute residual and the matrix $(X^{\rm T}X)^{-1}$ of the fit, where $X$ is the design matrix. A cell is used if it holds at least 200 stars, with at least 30 in each hemisphere. If every star moves with $v_z=A\,{\rm sign}(b)$ relative to the disc, then $S\approx A\langle\varpi/\cos b\rangle/K$, so $S>0$ for expansion and $S<0$ for compression. The term in $W_\odot$ is even in $b$ and is absorbed by $\beta_c\cos b$, so changing $W_\odot$ leaves $S$ unchanged (Extended Data Fig.~5). The $\cos b$ term is about $-0.04$\,mas\,yr$^{-1}$ in every region, including the control; it absorbs any error in the solar vertical velocity, vertical motions that are the same on both sides of the plane, and the all-sky proper-motion zero-point\cite{CantatGaudin2021}, and it does not enter $S$.

We keep $S$ in mas\,yr$^{-1}$. Converting it into a velocity requires the distance, and a parallax error $\delta\varpi$ then enters through the term $v_r\tan b\,\delta\varpi/K$, which changes sign with $b$ in the same way as a vertical motion. The distance bins are also defined by the observed parallax. In the model without vertical motion, the noise and this selection give regional values between $-0.024$ and $+0.016$\,mas\,yr$^{-1}$ (Table~1). We therefore compare the data only with mock observations that pass through the same selection and statistic.

\subsection{Cells and regions.}
Cells are defined by $2^\circ$ bins in longitude, from $-50^\circ$ to $+50^\circ$, and 11 bins in parallax distance $1/(\varpi+\varpi_0)$, with edges at 0.5, 1, 2, 3, 4, 5, 6, 7, 8, 9, 11 and 15\,kpc. The distance bins are grouped into 0.5--3, 3--6 and 6--15\,kpc. Fig.~1 uses 3--15\,kpc, Table~1 uses 3--6 and 6--15\,kpc, and Extended Data Fig.~3 also shows 0.5--3\,kpc. For the latitude tests, the sample is also split into two bands ($2^\circ\leq|b|<6^\circ$ and $6^\circ\leq|b|<10^\circ$) and into four bands $2^\circ$ wide. The regions are the near side ($l=4^\circ$ to $16^\circ$), the centre ($-4^\circ$ to $4^\circ$), the far side ($-20^\circ$ to $-8^\circ$) and the control ($|l|\geq24^\circ$). The longitudes between them ($-24^\circ$ to $-20^\circ$, $-8^\circ$ to $-4^\circ$ and $16^\circ$ to $24^\circ$) belong to no region. The four regions and the two distance groups were fixed, together with the criteria below, before the comparisons with the model reported here. Cells are combined into longitude points and regional values with inverse-variance weights.

Of the 5,160,487 stars that fall into cells, 5,143,370 (99.67\%) remain after the $5\sigma$ rejection. The 550 valid cells contain 1,158--25,066 stars each (median 9,441), and none of the 400 cells at 3--15\,kpc has fewer than 1,000 stars. Each longitude point in Fig.~1 combines 28,000--117,000 stars, and the near-side, central and far-side regions at 6--15\,kpc contain 291,000, 211,000 and 161,000 stars. The median scatter of the single-star residuals is about 1.7\,mas\,yr$^{-1}$, about 30 times the signal.

\subsection{N-body model.}
We generated self-consistent initial conditions for the Milky Way model of Tepper-Garc\'ia et al.\cite{TepperGarcia2021} with AGAMA\cite{Vasiliev2019}. The disc is exponential, with central surface density $1.1\times10^9\,M_\odot$\,kpc$^{-2}$, scale length 2.5\,kpc, a sech$^2$ vertical profile with scale height 0.3\,kpc, a quasi-isothermal distribution function\cite{Binney2010} with $\sigma_{R,0}=70$\,km\,s$^{-1}$ and $R_\sigma=5$\,kpc, and a total mass of $4.3\times10^{10}\,M_\odot$. The bulge is a Hernquist spheroid\cite{Hernquist1990} with a mass of $1.3\times10^{10}\,M_\odot$ and a scale radius of 0.6\,kpc, truncated at 2\,kpc. The dark halo follows a Navarro--Frenk--White profile\cite{NavarroFrenkWhite1996} with $\rho_0=9\times10^6\,M_\odot$\,kpc$^{-3}$ and $r_s=19$\,kpc, truncated at 250\,kpc, with a total mass of $1.17\times10^{12}\,M_\odot$; the value of $9\times10^9\,M_\odot$\,kpc$^{-3}$ printed in the original table is inconsistent with the total mass given there. The disc, bulge and halo contain $1.5\times10^6$, $3\times10^5$ and $3\times10^6$ particles, with softening lengths of 40\,pc for the disc and bulge and 100\,pc for the halo. We integrated the model for 6\,Gyr with GADGET-4\cite{Springel2021}, writing a snapshot every 25\,Myr.

A bar forms between about 1.0 and 1.5\,Gyr, as the $m=2$ Fourier amplitude of the disc, $A_2$, rises from 0.06 to 0.26, and by about 2\,Gyr the bar has thickened into a boxy/peanut shape. At 2.9\,Gyr, $A_2=0.30$, the bar half-length is 3.9\,kpc and the pattern speed is 38.5\,km\,s$^{-1}$\,kpc$^{-1}$. Each epoch used here combines eight snapshots spanning 175\,Myr: 2.5\,Gyr (41.9\,km\,s$^{-1}$\,kpc$^{-1}$), 2.9\,Gyr (38.5\,km\,s$^{-1}$\,kpc$^{-1}$; main comparison) and 3.5\,Gyr (35.9\,km\,s$^{-1}$\,kpc$^{-1}$), with 4.5\,Gyr (30.6\,km\,s$^{-1}$\,kpc$^{-1}$) and 5.9\,Gyr (27.3\,km\,s$^{-1}$\,kpc$^{-1}$) used only to test sensitivity. Near 2.4\,Gyr, the pattern speed of our run, about 43\,km\,s$^{-1}$\,kpc$^{-1}$, agrees with that of the original simulation\cite{TepperGarcia2021}, but afterwards it declines faster than in that simulation, by 5.4\,km\,s$^{-1}$\,kpc$^{-1}$ per Gyr between 2.4 and 4.3\,Gyr. The lower particle number of our run may contribute to this difference\cite{Athanassoula2003,Dubinski2009,Fujii2018,Ludlow2021}. We therefore select the main epoch by its pattern speed.

The public snapshot of the original simulation\cite{TepperGarcia2021} (Extended Data Fig.~8) is a separate, higher-resolution integration of the same model at 4.28\,Gyr; its halo contains 1.54\% fewer particles than the nominal number. It passes through the same mock observations. We did not measure its pattern speed.

\subsection{Mock observations.}
Every epoch and bar angle, and the no-bar control described below, pass through the same procedure. We place the Sun in the model at a Galactocentric radius of 8.0\,kpc and a height of 20.8\,pc above the plane\cite{BennettBovy2019}, with the bar at an angle of $27^\circ$ to the line from the Sun to the Galactic Centre\cite{Wegg2013,Portail2017,BlandHawthorn2016}; we also use angles from $15^\circ$ to $45^\circ$. These values only place the Sun and the bar in the model; the statistic itself uses none of them. The solar velocity is $(11.1,\,12.24+v_c,\,7.78)$\,km\,s$^{-1}$, where $v_c$ is the mean azimuthal velocity of cold disc particles at radii of 7.5--8.5\,kpc in the model; the statistic uses the same $W_\odot$. We select particles at $|l|<51^\circ$, $2^\circ\leq|b|<10^\circ$ and true distances of 0.2--25\,kpc. Each particle receives a parallax $\varpi=1/d-\varpi_0+\sigma_\varpi\,\mathcal{N}(0,1)$, where $d$ is its true distance and $\sigma_\varpi$ is drawn from data stars in the same $2^\circ$ longitude bin and latitude band, and only $\varpi>0$ is kept. We weight the particles so that their numbers match those of the data in each cell of longitude, latitude band (two bands) and parallax-distance bin, and draw $6\times10^6$ mock stars per epoch with replacement. The weights do not distinguish the two hemispheres. In the cells at 3--15\,kpc, the median ratio of stars north and south of the plane is 0.96 on the near side and at the centre, 0.74 on the far side and 0.62 in the control in the data, and 0.90--0.94 in all four regions in the model. Uncertainties in $\mu_b$ and radial velocity are drawn from data stars in the same cell and added as Gaussian noise. For the model without vertical motion, we use the same particles and the same noise but remove each particle's own vertical velocity from $\mu_b$ and the radial velocity. The mock stars then pass through the same statistic. Cells with fewer than 500 particles, counted over the eight snapshots of the epoch (one for the public snapshot), are marked as model-limited and excluded from comparisons.

\subsection{Uncertainties.}
The formal uncertainty of $S$, computed from 1.4826 times the median absolute residual, underestimates the scatter of the data, partly because the residuals have heavier tails than a Gaussian. To calibrate it, we split the stars at random into two halves and measure $S$ in each half of every cell. If the formal uncertainties were correct, the differences between the halves, divided by their formal uncertainties, would have a standard deviation of 1; over ten random splits the median is 1.31, and we multiply the statistical uncertainties of the data by this factor.

In many cells of the model the mock stars outnumber the particles, so some particles are drawn more than once. We therefore increase the formal uncertainty $\sigma_S$ of a model value to $\sigma_{\rm eff}=\sigma_S\,[\max(1,n_{\rm draw}/n_{\rm part})]^{1/2}$, where $n_{\rm draw}$ is the number of mock stars and $n_{\rm part}$ the number of particles in the cell, counted over the snapshots of the epoch. To calibrate $\sigma_{\rm eff}$, we observe the first and the last four of the eight snapshots of the 2.9\,Gyr epoch separately. The differences between the two halves, divided by $\sigma_{\rm eff}$, have a median absolute value of 1.12, against 0.674 for a Gaussian, so we multiply $\sigma_{\rm eff}$ by 1.66; this factor also includes changes of the model within the epoch. For the public snapshot, whose halves are alternate particles of the same snapshot, the factor is 1.32. All figures, Table~1 and the criteria use these factors.

Systematic terms are added in quadrature to the statistical uncertainty of each regional value. Every comparison includes three such terms. The first is the larger change of the model value when the mock parallaxes carry a zero-point of 0.015 or 0.035\,mas while the measurement still corrects for 0.025\,mas. The second is a projection term. The value of the model without vertical motion arises from noise and selection acting on the radial-velocity term of the subtraction, and therefore scales with the part of $S$ that this term removes; the projection term is this value multiplied by the fractional difference between data and model in that part. The third is the spatially correlated error of Gaia proper motions, described below. Comparisons with the bar model also include the standard deviation of the model among the 2.5, 2.9 and 3.5\,Gyr epochs.

Gaia proper motions also contain systematic errors that are correlated across the sky\cite{Lindegren2021astrometry,VasilievBaumgardt2021}. A rotation of the reference frame, such as the spin of the bright-star frame\cite{CantatGaudin2021}, changes $\mu_b$ by an amount that does not depend on $b$, and the $\cos b$ term absorbs it. A uniform shift of $\mu_b$ changes the regional values in Table~1 by at most 1\% of the shift. For errors that vary across the sky we use the covariance of the proper motions of two stars separated by $\theta$ derived by Vasiliev and Baumgardt\cite{VasilievBaumgardt2021}, $V_\mu(\theta)=400/(1+\theta/0.3^\circ)+300\exp(-\theta/12^\circ)\,\mu{\rm as}^2\,{\rm yr}^{-2}$. With the clipping fixed, each regional value is a linear combination of the residuals of its stars, so its variance is the sum of $V_\mu$ over all pairs of stars, weighted by their contributions; we evaluate it with the stars grouped in equal-area sky pixels of $0.23^\circ$. This gives 0.004--0.009\,mas\,yr$^{-1}$ for the regional values in Table~1. The covariance was derived from cluster stars fainter than $G=13$ and from quasars, and our sample includes brighter stars; doubling it would multiply these values by 1.41, and a correlation length of $6^\circ$ or $24^\circ$ instead of $12^\circ$ multiplies them by 0.87--1.03. We add this term in quadrature to the regional uncertainties in Table~1 and in the predefined criteria; the uncertainties of single cells and longitude bins in the figures do not include it. Without this term, the $z$ values relative to no vertical motion would be $-5.15$ and $-6.52$ on the near side, $9.42$ at the centre and $3.61$ on the far side.

Correlations between longitude curves use the 24 bins at $|l|<24^\circ$ and 3--15\,kpc. The amplitude ratio $s$ of data $d$ to model $m$ minimizes $\chi^2(s)=\sum(d-sm)^2/(e^2+s^2\sigma^2)$, where $e$ and $\sigma$ are the data and model uncertainties, and its 68\% interval follows from $\Delta\chi^2=1$.

\subsection{Predefined criteria.}
We fixed the following criteria and thresholds before the comparisons of data and model reported here, and did not relax any threshold afterwards. Two terms of the uncertainties, the spatially correlated errors of Gaia proper motions and the projection term in the comparisons with the model without vertical motion (Uncertainties), were added after the first comparisons; they only enlarge the uncertainties and change none of the outcomes below. Detection: the data are compared with the model without vertical motion in each region and distance group; $z\geq5$ counts as a detection and $3\leq z<5$ as marginal, and compression is reported with its signed $z$. Agreement on the near side and at the centre: the data are compared with the 2.9\,Gyr model in the 3--6 and 6--15\,kpc groups of these two regions; agreement requires $|z|<3$ and the same sign in all four. Far-side discrepancy: the data are compared with the model in the far-side 6--15\,kpc group at 2.5, 2.9 and 3.5\,Gyr and at bar angles of $20^\circ$, $25^\circ$, $30^\circ$ and $35^\circ$; a robust discrepancy requires $|z|\geq3$ in all seven comparisons. Location of the far-side residual: for parallax-distance bins in which the far side differs from the model by $|z|\geq3$, the true distances of the mock stars show whether the signal is concentrated in distance, whether it lies beyond the point where the sightline passes closest to the Galactic Centre, and whether it is associated with the bar. Sky-fixed proper-motion errors: in each region we compare four descriptions of the residual by their Bayesian information criterion (BIC)\cite{Schwarz1978}, namely zero, one constant, one constant per longitude bin and one constant per parallax-distance bin. If the BIC of the per-distance description is lower than each of the other three by at least 6, the residual is classed as distance-dependent; otherwise, if the BIC of the constant or the per-longitude description is lower than that of zero by at least 6, the residual is classed as possibly sky-fixed; otherwise it is consistent with zero. We apply this test both to the data minus the 2.9\,Gyr model and to the data minus the model without vertical motion.

\subsection{Outcomes of the criteria and further tests.}
Detection: the central expansion at 6--15\,kpc is a detection, the far side at 6--15\,kpc is marginal and the control is consistent with zero (Table~1). The detection labels apply to positive $z$; the near-side compression is reported with its signed $z$, $-4.28$ at 3--6\,kpc and $-4.43$ at 6--15\,kpc, both between the marginal and detection thresholds in absolute value. Agreement on the near side and at the centre is formally not met, because in the central 3--6\,kpc group the data and the model have opposite signs, although $|z|=0.67$; the Results explain this difference. The far-side discrepancy is not robust by this criterion, because $|z|\geq3$ in only two of the seven comparisons: $z=2.66$, 2.44 and 3.37 at 2.5, 2.9 and 3.5\,Gyr, and 1.77, 2.29, 2.58 and 3.09 at bar angles of $20^\circ$, $25^\circ$, $30^\circ$ and $35^\circ$. Supplementary Fig.~1 shows the data minus the model cell by cell. Only the 9--11\,kpc bin on the far side meets the condition of the location test; its mock stars are concentrated in true distance, lie beyond the point of closest approach to the Galactic Centre and are associated with the bar.

No-bar control: the same simulation at 0.40--0.57\,Gyr (snapshots 16--23, $A_2<0.04$), before the bar forms, passes through the same mock observations; its correlation with the data is $-0.16$.

Far-side turnover: on the 6--15\,kpc longitude curve we start with the bin from $l=-6^\circ$ to $-4^\circ$ and move towards negative longitudes until we find three consecutive positive $2^\circ$ bins; the turnover is the centre of the first of these bins, the one closest to the Galactic Centre. We measured it for the data, for the model at a bar angle of $27^\circ$ at nine epochs between 2.3 and 3.9\,Gyr, spaced by 0.2\,Gyr and each combining eight snapshots, and for the 2.9\,Gyr model at bar angles from $15^\circ$ to $45^\circ$. The far-side longitude ranges quoted in the Results refer to bin centres. This definition was made after seeing the data, serves only as a description and is not one of the criteria.

Sky-fixed errors: relative to the 2.9\,Gyr model, the near side, the centre and the control are consistent with zero. On the far side a single constant offset is preferred (BIC 44, against 75 for zero, 62 for one value per distance bin and 61 for one value per longitude bin), so this residual could be sky-fixed. The far-side difference between data and model at 6--15\,kpc, 0.067\,mas\,yr$^{-1}$, is 8 times the uncertainty that the spatially correlated errors of Gaia proper motions give for this regional value (Uncertainties). Relative to the model without vertical motion, the near-side, central and far-side signals all depend on parallax distance; the control signal does not.

Alternative choices and subsamples: with the same stars, we changed the parallax zero-point to 0.015 and 0.035\,mas, set $W_\odot=7.25$\,km\,s$^{-1}$, removed the radial-velocity zero-point correction and clipped at $3\sigma$ and $8\sigma$. We also used subsamples with RUWE $<1.2$, radial-velocity uncertainty $<5$\,km\,s$^{-1}$ and relative parallax uncertainty $<15\%$, four magnitude ranges ($G<13$, 13--14, 14--15 and 15--16), three equal-size colour groups and three groups in the number of Gaia scans, formed at fixed sky position as described next. Every alternative choice keeps the signs of the full sample in the three signal regions at 6--15\,kpc. Among the subsamples, the only sign change is on the far side for $G<13$ (574 stars), where $S=-0.05\pm0.15$\,mas\,yr$^{-1}$ is consistent with both zero and the full sample (Extended Data Fig.~5).

Number of Gaia scans: we use the number of visibility periods in the astrometric solution of each star, which depends on its position on the sky through the scanning law. Over a whole region, groups of stars with different numbers of visibility periods therefore also lie at different positions, and such groups differ by up to $z=+9.04$ (control, 3--6\,kpc). We therefore divide the stars into three equal groups by this number within each $2^\circ\times2^\circ$ field, hemisphere and parallax-distance bin, and compare each group with the other stars of the same cells (Extended Data Fig.~5 shows the groups at 6--15\,kpc). No group then differs by $|z|\geq3$ in any region and distance group; the largest difference is $z=-2.31$, on the far side at 3--6\,kpc. Before the test we required that $z$ values from randomly shuffled groups have a standard deviation between 0.8 and 1.3, so that $z$ can be read as a significance. The shuffled groups give 1.319, slightly above this range, which means that the $z$ values overstate the differences; the result that no group differs by $|z|\geq3$ is therefore unaffected, and we find no dependence on the number of scans.

Latitude: all four $2^\circ$ latitude bands between $|b|=2^\circ$ and $10^\circ$ show the same sequence of signs along longitude (Extended Data Fig.~4). The change from $2^\circ\leq|b|<6^\circ$ to $6^\circ\leq|b|<10^\circ$ differs between data and model by at most $0.69\sigma$ in any region. In single $2^\circ$ bands, the largest difference between data and model is on the far side at $4^\circ\leq|b|<6^\circ$, $2.5\sigma$, which reflects the overall far-side offset.

\subsection{Parametric velocity fields.}
We also tried to describe the data with parametric fields of vertical velocity in the bar frame. Mock stars without vertical motion were used to build a response matrix that converts each field into $S$ in each cell, and the fields were compared on the same cells (Supplementary Figs.~2 and 3). When the selected field was injected back into the mock observations, the measured values in two region groups differed from the linear prediction by more than 0.2 times their uncertainty (by up to 0.004\,mas\,yr$^{-1}$), so the response is not linear for this field. We therefore do not adopt these fits and do not use them to correct the N-body comparison.

\subsection{Data availability.}
Gaia DR3 data are publicly available from the Gaia Archive (\texttt{https://gea.esac.esa.int/archive/}). The processed catalogue, the cell measurements and the mock tables are available at Zenodo\cite{Wan2026data} (\texttt{https://doi.org/10.5281/zenodo.24523014}). The public snapshot of the original simulation is available from the source given by Tepper-Garc\'ia et al.\cite{TepperGarcia2021}.

\subsection{Code availability.}
The code for the measurement and figures (\texttt{obs\_breathing.py}) and for the N-body initial conditions (\texttt{nbody\_ics.py}) is available from the corresponding author upon reasonable request. The analysis uses AGAMA\cite{Vasiliev2019}, GADGET-4\cite{Springel2021}, Astropy\cite{Astropy2013,Astropy2018,Astropy2022}, NumPy\cite{Harris2020}, SciPy\cite{Virtanen2020}, Matplotlib\cite{Hunter2007} and pandas\cite{McKinney2010}.

\subsection{Use of generative AI.}
Parts of the analysis code were drafted, and the manuscript was edited for language, with the assistance of Cursor (Anysphere). Z.W. reviewed the code and the text and takes responsibility for them.

\end{methods}

\bibliographystyle{naturemag}
\bibliography{references}

\begin{addendum}
 \item[Acknowledgements] This work is supported by the National Key R\&D Program of China No. 2025YFF0511003. This work has made use of data from the European Space Agency (ESA) mission Gaia (\texttt{https://www.cosmos.esa.int/gaia}), processed by the Gaia Data Processing and Analysis Consortium (DPAC, \texttt{https://www.cosmos.esa.int/web/gaia/dpac/consortium}). Funding for the DPAC has been provided by national institutions, in particular the institutions participating in the Gaia Multilateral Agreement.
 \item[Author contributions] Z.W. designed the study, carried out the analysis and wrote the manuscript. L.F. contributed to the discussion and provided the funding.
 \item[Competing interests] The authors declares no competing interests.
 \item[Correspondence] Correspondence and requests for materials should be addressed to Zhen Wan (zhen\_wan@ustc.edu.cn).
\end{addendum}

\clearpage
\spacing{1}

\noindent\includegraphics[width=\textwidth]{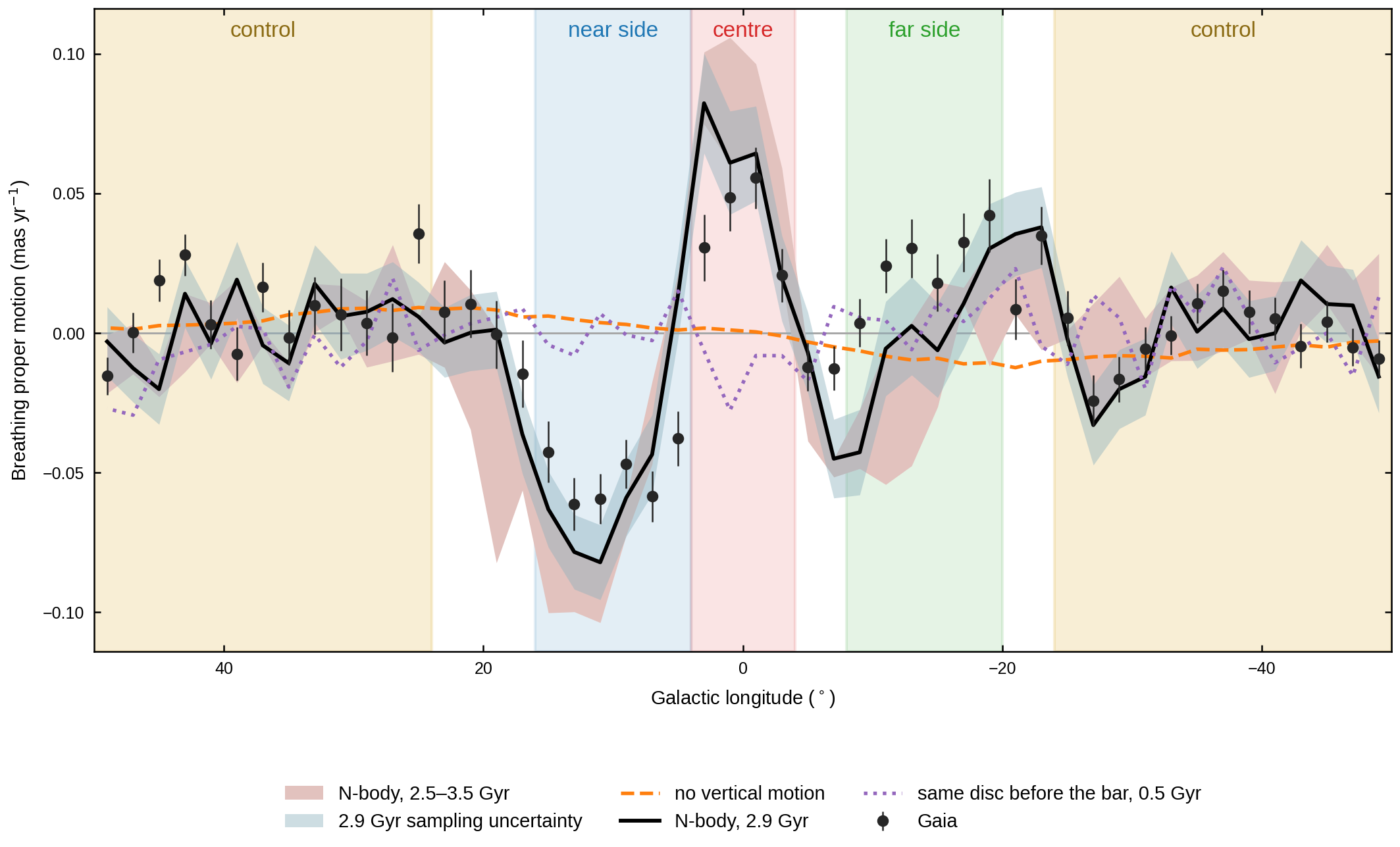}

\noindent\textbf{Fig.~1 $|$ Breathing proper motion along Galactic longitude.} Points are Gaia DR3 values of the breathing proper motion $S$ in $2^\circ$ longitude bins, combining parallax distances of 3--15\,kpc (28,000--117,000 stars per bin); error bars are $1\sigma$. The orange dashed line is the N-body model observed with the same selection and noise after the vertical velocity of each particle is removed. The black line is the N-body bar at 2.9\,Gyr (pattern speed 38.5\,km\,s$^{-1}$\,kpc$^{-1}$), observed in the same way as the data; the blue-grey band is its $1\sigma$ sampling uncertainty. The rose band spans the model at 2.5, 2.9 and 3.5\,Gyr. The purple dotted line is the same disc at 0.5\,Gyr, before the bar formed. Shaded intervals mark the near side ($l=4^\circ$ to $16^\circ$, blue), the centre ($-4^\circ$ to $4^\circ$, red), the far side ($-20^\circ$ to $-8^\circ$, green) and the control longitudes ($|l|\geq24^\circ$, gold). Positive values indicate expansion and negative values compression. Longitude increases to the left.

\clearpage

\noindent\includegraphics[width=\textwidth]{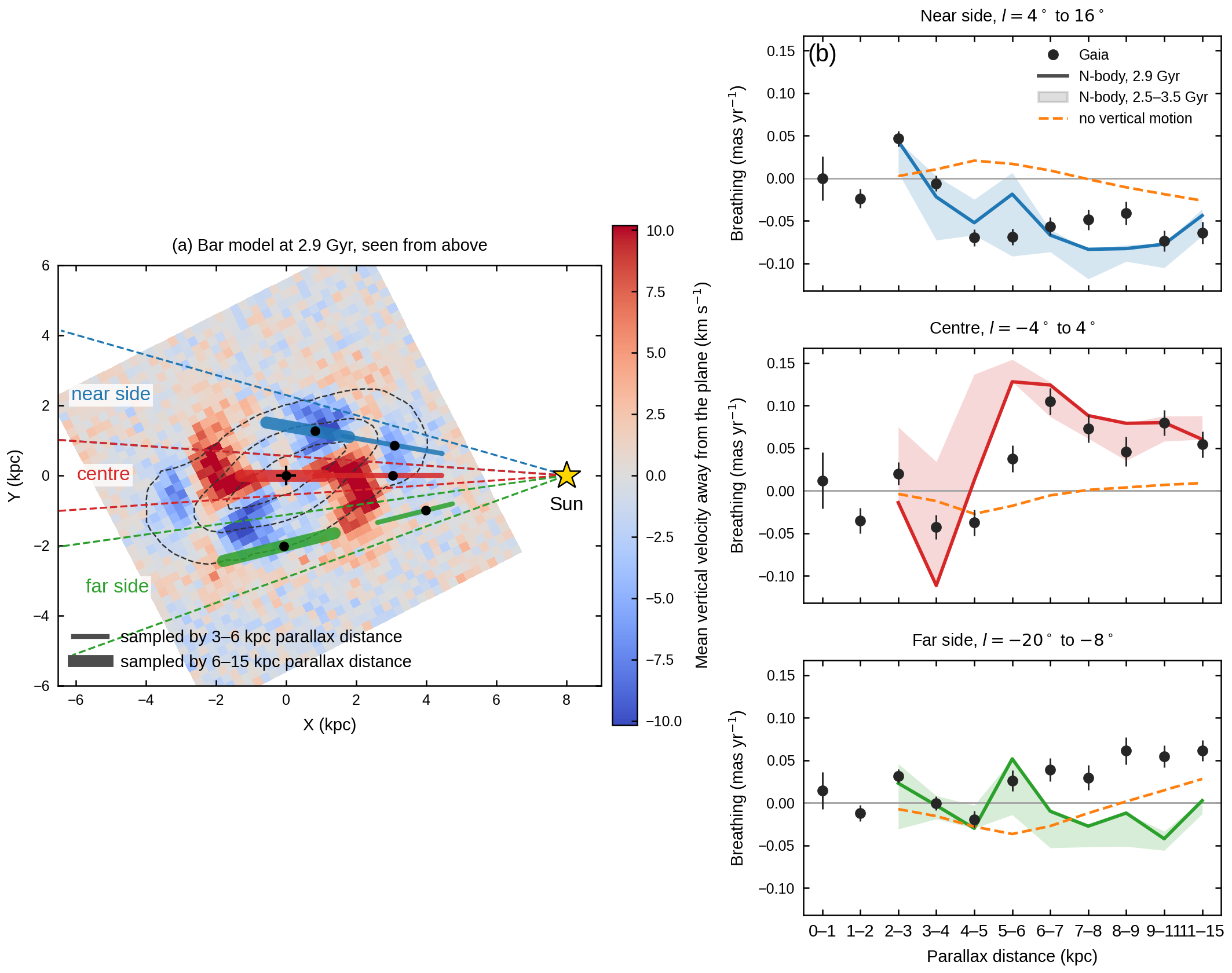}

\noindent\textbf{Fig.~2 $|$ Where the breathing signal arises along the line of sight.} \textbf{a}, The N-body bar at 2.9\,Gyr seen from above, with the bar at $27^\circ$ to the line from the Sun to the Galactic Centre. Colour is the mean vertical velocity away from the plane in km\,s$^{-1}$ (positive for expansion, negative for compression), averaged over the eight snapshots of this epoch; contours trace the stellar surface density. The star marks the Sun and the cross the Galactic Centre. Dashed lines bound the near-side (blue), central (red) and far-side (green) sightlines. On each sightline, the thin and thick segments span the 16th to 84th percentiles of the true distances of mock stars in the 3--6 and 6--15\,kpc parallax-distance groups, weighted by their contribution to $S$ ($\propto|\sin b|/d$); black dots mark the weighted medians. \textbf{b}, Breathing proper motion against parallax distance for the near side, centre and far side. Points are Gaia with $1\sigma$ error bars, solid lines are the 2.9\,Gyr model, shaded bands span the model at 2.5--3.5\,Gyr and orange dashed lines are the model without vertical motion. Model values are not shown below 2\,kpc, where cells contain fewer than 500 model particles.

\clearpage

\noindent\textbf{Table 1 $|$ Regional breathing proper motions.}

\vspace{6pt}
\noindent
{\small\setlength{\tabcolsep}{4pt}%
\begin{tabular}{llrrrrrr}
\toprule
 & & \multicolumn{3}{c}{Gaia versus no vertical motion} & \multicolumn{3}{c}{Gaia versus 2.9\,Gyr model} \\
\cmidrule(lr){3-5}\cmidrule(lr){6-8}
Region & Distance & Gaia & No vertical motion & $z$ & Gaia & Model & $z$ \\
\midrule
Near side & 3--6\,kpc & $-0.0465$ & $+0.0159$ & $-4.28$ & $-0.0591$ & $-0.0315$ & $-0.99$ \\
Near side & 6--15\,kpc & $-0.0570$ & $-0.0075$ & $-4.43$ & $-0.0589$ & $-0.0720$ & $+0.53$ \\
Centre & 3--6\,kpc & $-0.0163$ & $-0.0188$ & $+0.17$ & $-0.0135$ & $+0.0175$ & $-0.67$ \\
Centre & 6--15\,kpc & $+0.0728$ & $+0.0030$ & $+5.98$ & $+0.0708$ & $+0.0832$ & $-0.57$ \\
Far side & 3--6\,kpc & $-0.0007$ & $-0.0239$ & $+1.05$ & $+0.0051$ & $+0.0109$ & $-0.16$ \\
Far side & 6--15\,kpc & $+0.0498$ & $+0.0032$ & $+3.06$ & $+0.0544$ & $-0.0124$ & $+2.44$ \\
Control & 3--6\,kpc & $+0.0056$ & $-0.0024$ & $+1.22$ & $+0.0042$ & $-0.0051$ & $+0.92$ \\
Control & 6--15\,kpc & $-0.0016$ & $+0.0003$ & $-0.31$ & $-0.0016$ & $+0.0007$ & $-0.33$ \\
\bottomrule
\end{tabular}}

\vspace{6pt}
\noindent Values of $S$ are in mas\,yr$^{-1}$; distances are parallax distances. Columns 3--5 compare Gaia with the N-body model observed without vertical motion, and columns 6--8 with the 2.9\,Gyr model. Both comparisons use the same cells, each weighted by the inverse square of the combined uncertainty of data and model; the 2.9\,Gyr model has the larger sampling uncertainty, so the two Gaia columns differ by up to 0.013\,mas\,yr$^{-1}$. Each $z$ is the difference between the two preceding columns divided by the uncertainty of that difference, with its sign kept; the text quotes these values as $\sigma$. Uncertainties combine the scaled statistical errors of data and model with the systematic terms described in Methods, including the spatially correlated errors of Gaia proper motions.

\clearpage

\noindent\includegraphics[width=\textwidth]{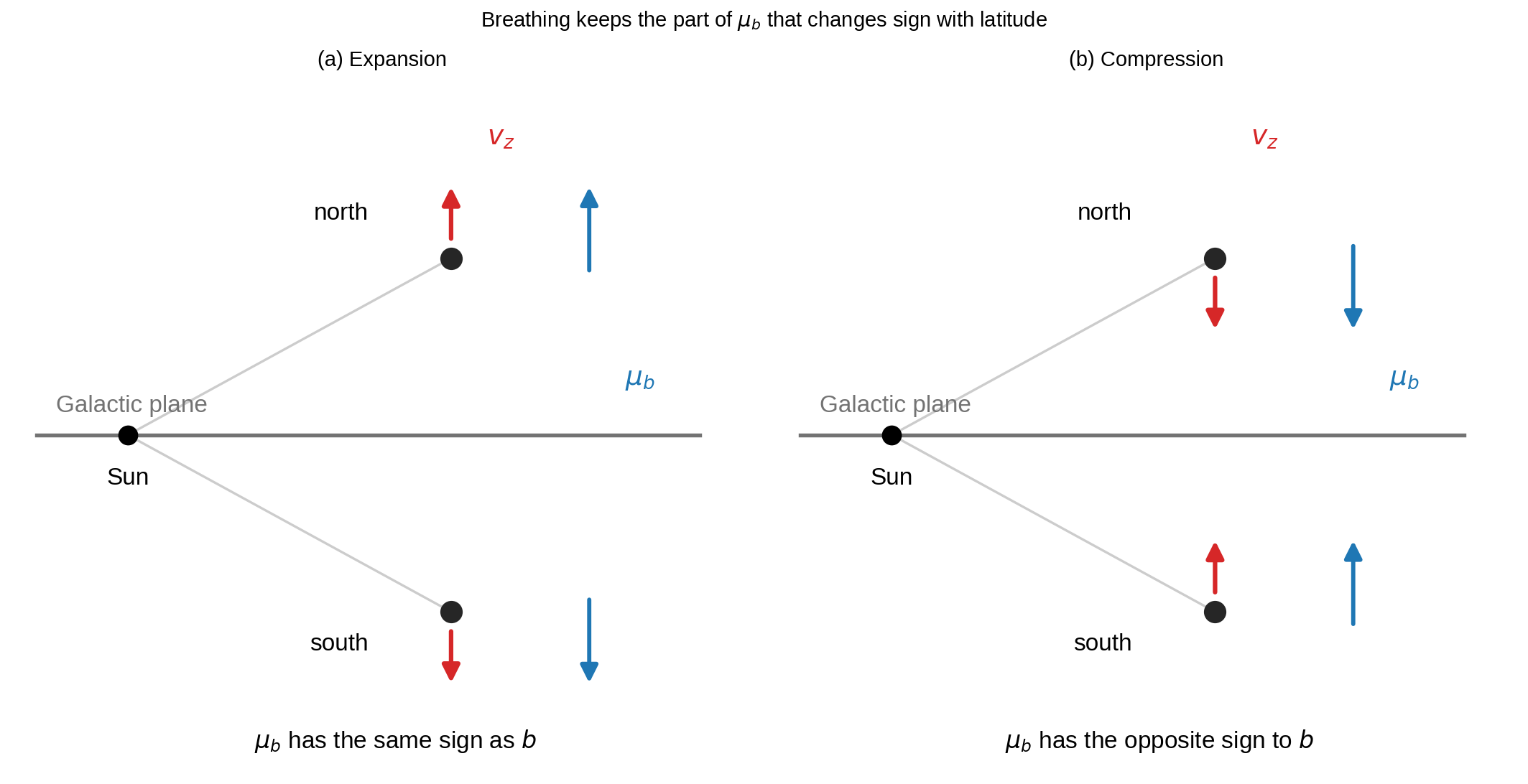}

\noindent\textbf{Extended Data Fig.~1 $|$ How vertical breathing appears in proper motion.} \textbf{a}, Expansion: stars north and south of the plane move away from it, and $\mu_b$ has the same sign as Galactic latitude $b$. \textbf{b}, Compression: stars move towards the plane, and $\mu_b$ has the opposite sign to $b$. Red arrows show the vertical velocity $v_z$ and blue arrows the resulting $\mu_b$. The breathing statistic keeps the part of $\mu_b$ that changes sign with latitude.

\clearpage

\noindent\includegraphics[width=\textwidth]{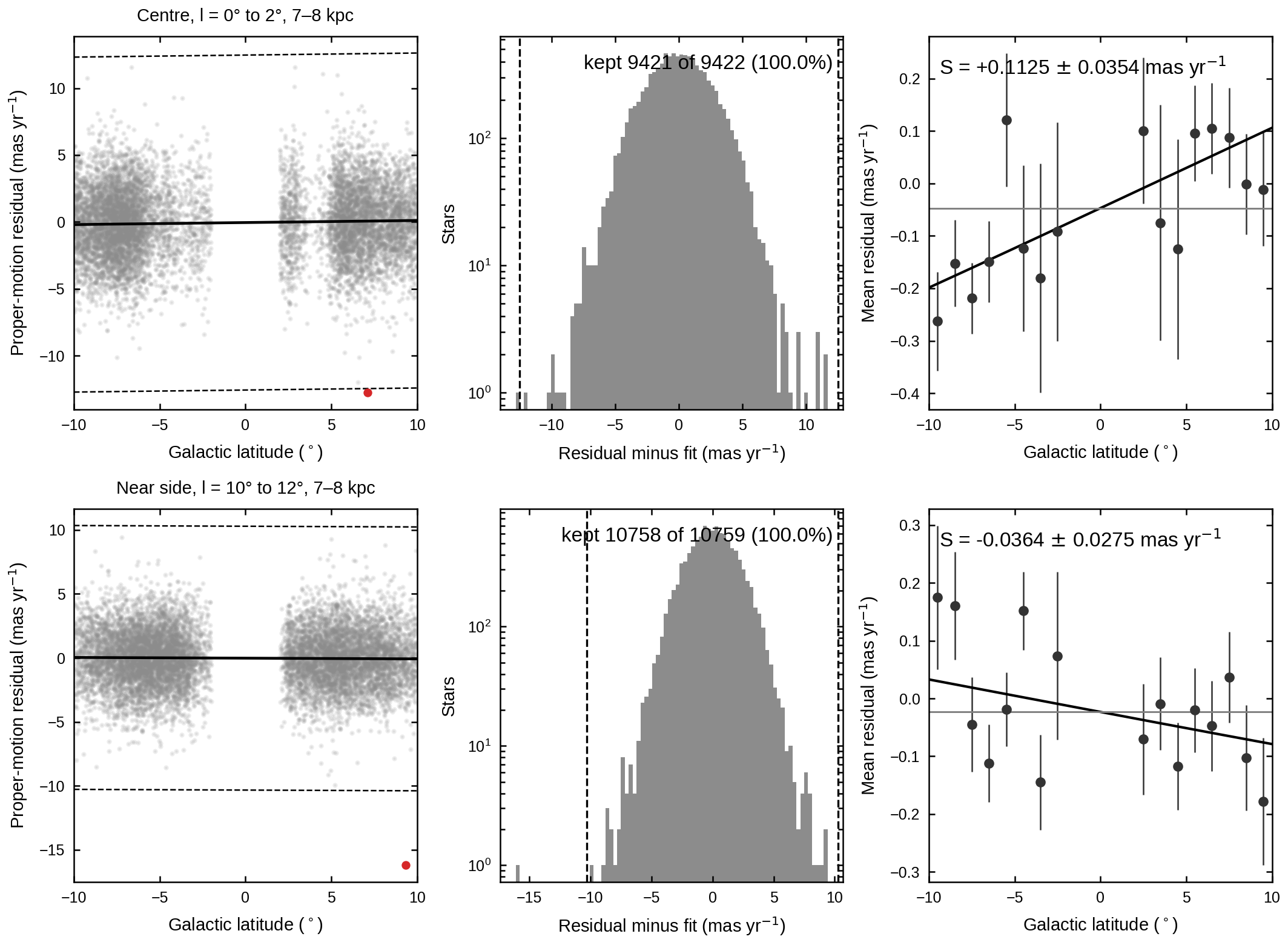}

\noindent\textbf{Extended Data Fig.~2 $|$ Measurement in two example cells.} Top row: centre, $l=0^\circ$ to $2^\circ$, parallax distance 7--8\,kpc. Bottom row: near side, $l=10^\circ$ to $12^\circ$, 7--8\,kpc. Left: proper-motion residual against Galactic latitude; grey points are kept and red points are rejected by the $5\sigma$ clip; the black curve is the fit in $\sin b$ and $\cos b$, and dashed curves mark the clip bounds. Middle: distribution of the residual minus the fit, with counts on a logarithmic axis and the clip bounds dashed; one star is rejected in each cell. Right: mean residual in $1^\circ$ latitude bins with standard errors, the same fit, and the $\cos b$ term (grey horizontal line); the value of $S$ and its $1\sigma$ uncertainty are given in each panel. Residuals are in mas\,yr$^{-1}$.

\clearpage

\noindent\includegraphics[width=\textwidth]{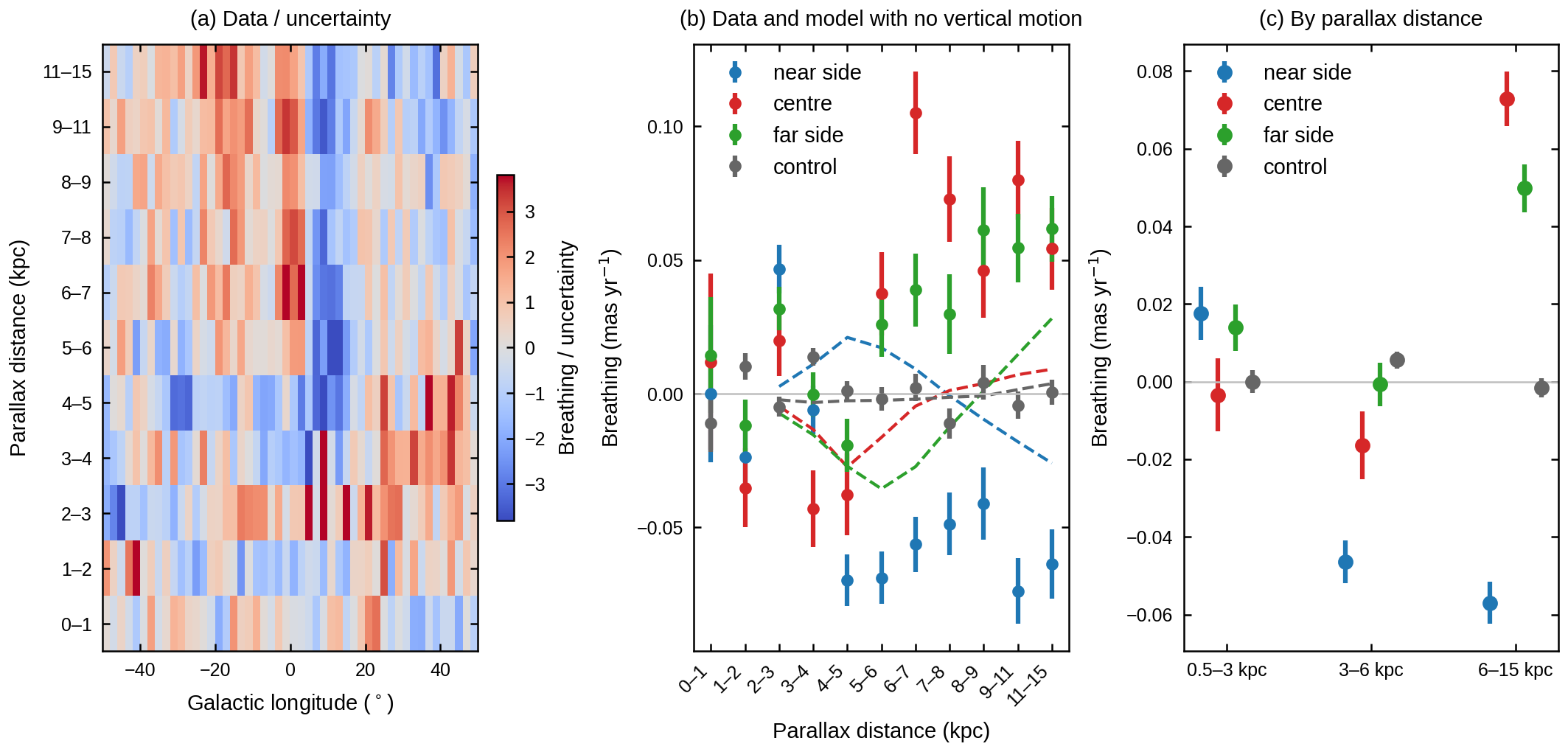}

\noindent\textbf{Extended Data Fig.~3 $|$ Breathing across longitude and parallax distance.} \textbf{a}, Breathing proper motion divided by its uncertainty in each cell of $2^\circ$ longitude and parallax distance. \textbf{b}, Breathing proper motion against parallax distance for the near side, centre, far side and control; dashed lines in matching colours are the model without vertical motion, shown from 2\,kpc outwards because closer cells contain fewer than 500 model particles. \textbf{c}, The same four regions in the 0.5--3, 3--6 and 6--15\,kpc parallax-distance groups. Breathing proper motions are in mas\,yr$^{-1}$; error bars are $1\sigma$.

\clearpage

\noindent\includegraphics[width=\textwidth]{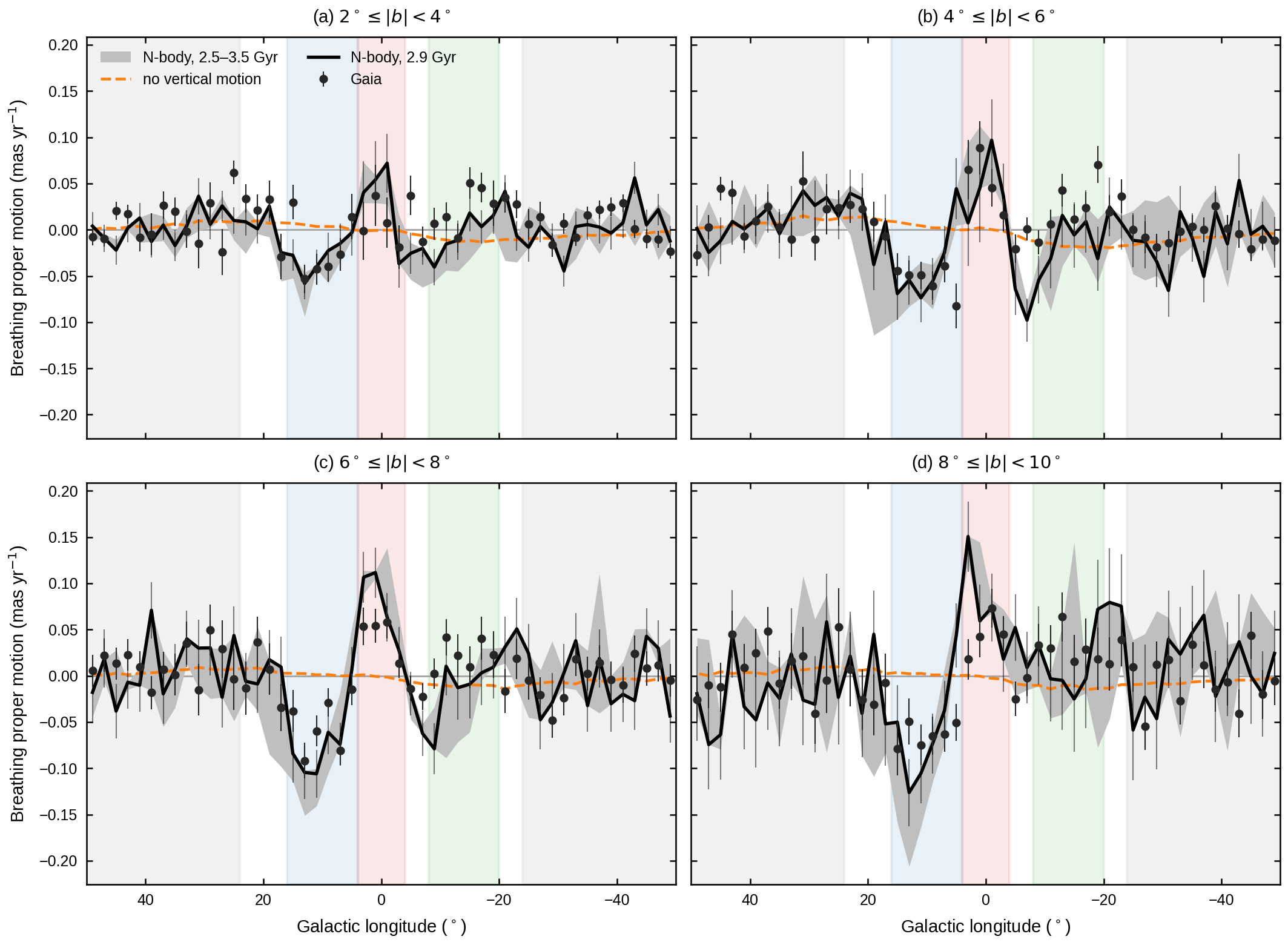}

\noindent\textbf{Extended Data Fig.~4 $|$ The longitude curve in four latitude bands.} Breathing proper motion against Galactic longitude at parallax distances of 3--15\,kpc in \textbf{a}, $2^\circ\leq|b|<4^\circ$; \textbf{b}, $4^\circ\leq|b|<6^\circ$; \textbf{c}, $6^\circ\leq|b|<8^\circ$; and \textbf{d}, $8^\circ\leq|b|<10^\circ$. Points are Gaia with $1\sigma$ error bars. The black line is the 2.9\,Gyr model, with its $1\sigma$ sampling uncertainty drawn as error bars; the grey band spans the model at 2.5--3.5\,Gyr; and the orange dashed line is the model without vertical motion. Shaded longitude intervals mark the same regions as in Fig.~1, with the control in grey. Values are in mas\,yr$^{-1}$. Longitude increases to the left.

\clearpage

\noindent\includegraphics[width=\textwidth]{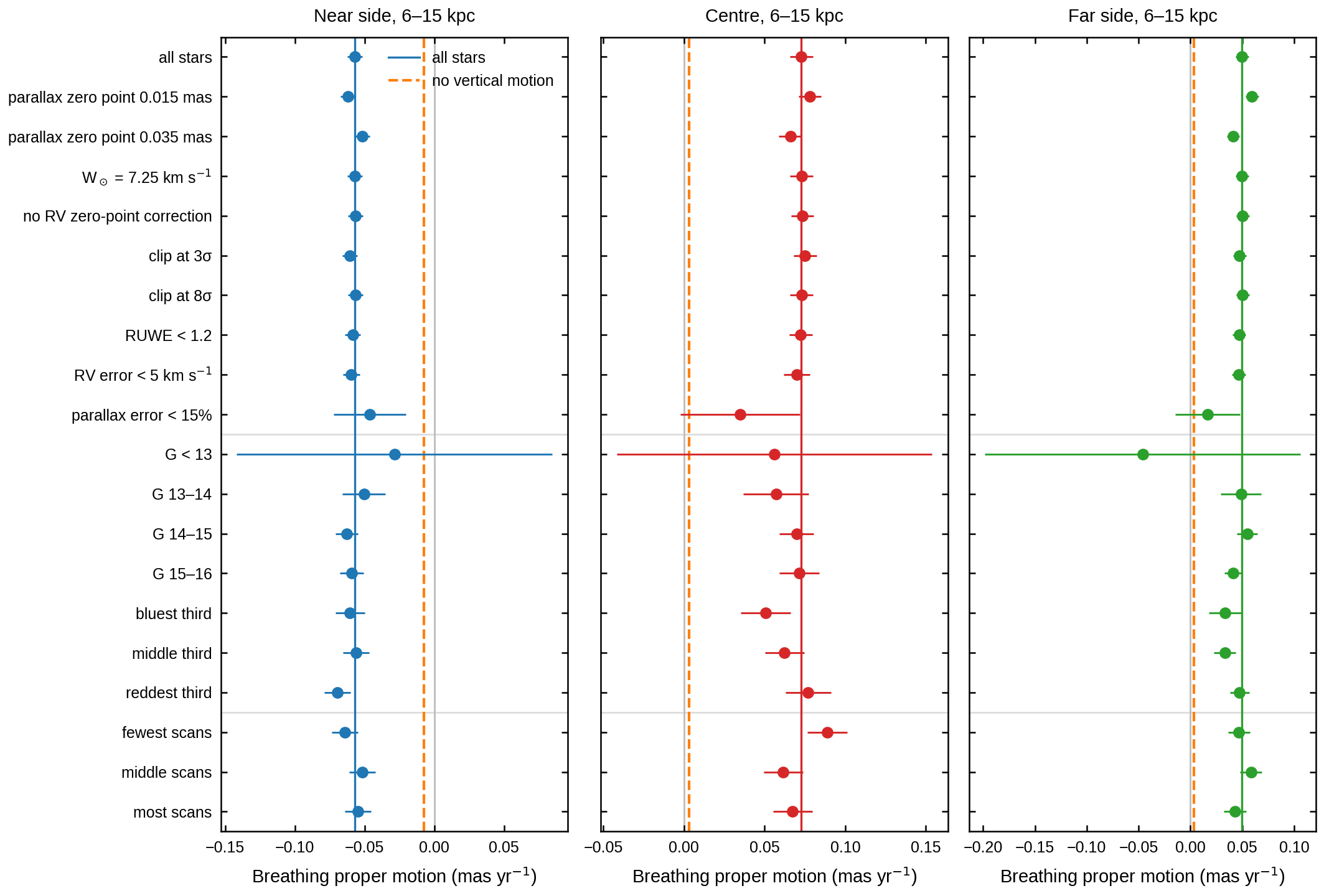}

\noindent\textbf{Extended Data Fig.~5 $|$ Alternative analysis choices.} Breathing proper motion at parallax distances of 6--15\,kpc for the near side, centre and far side. Each row is one choice: the full sample; parallax zero-points of 0.015 and 0.035\,mas; $W_\odot=7.25$\,km\,s$^{-1}$; no radial-velocity zero-point correction; clipping at $3\sigma$ and $8\sigma$; RUWE $<1.2$; radial-velocity uncertainty $<5$\,km\,s$^{-1}$; relative parallax uncertainty $<15\%$; four $G$-magnitude ranges; three colour groups; and three groups in the number of Gaia scans, formed at fixed sky position. Error bars are $1\sigma$. Solid vertical lines mark the full sample and dashed lines the model without vertical motion. Values are in mas\,yr$^{-1}$.

\clearpage

\noindent\includegraphics[width=0.75\textwidth]{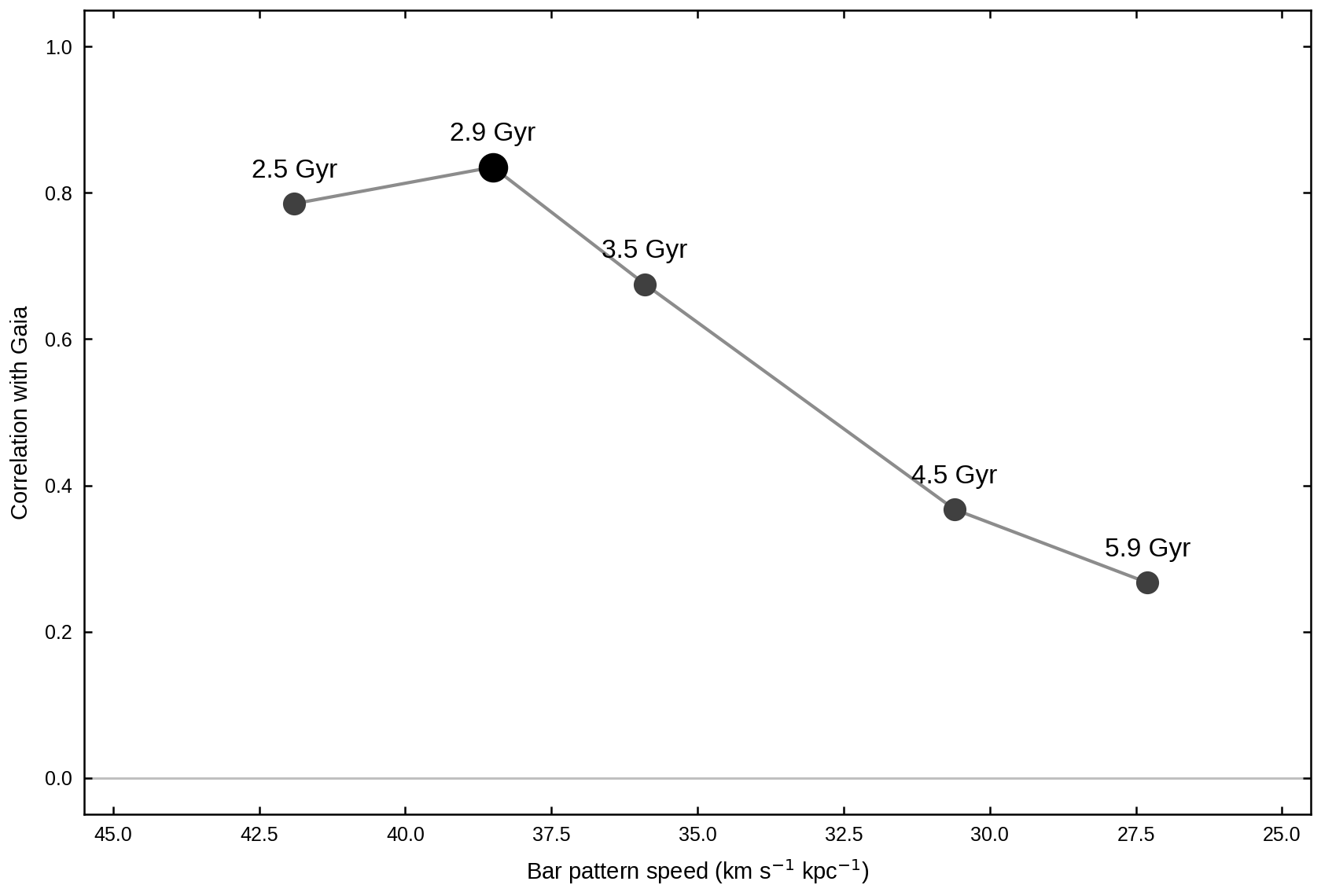}

\noindent\textbf{Extended Data Fig.~6 $|$ Agreement with the data at different pattern speeds.} Correlation between the model and Gaia longitude curves ($|l|<24^\circ$, 3--15\,kpc) at five epochs of the same simulation, plotted against the bar pattern speed at each epoch. The 2.9\,Gyr epoch (38.5\,km\,s$^{-1}$\,kpc$^{-1}$), used for the main comparison, is shown in black. Pattern speed decreases to the right.

\clearpage

\noindent\includegraphics[width=\textwidth]{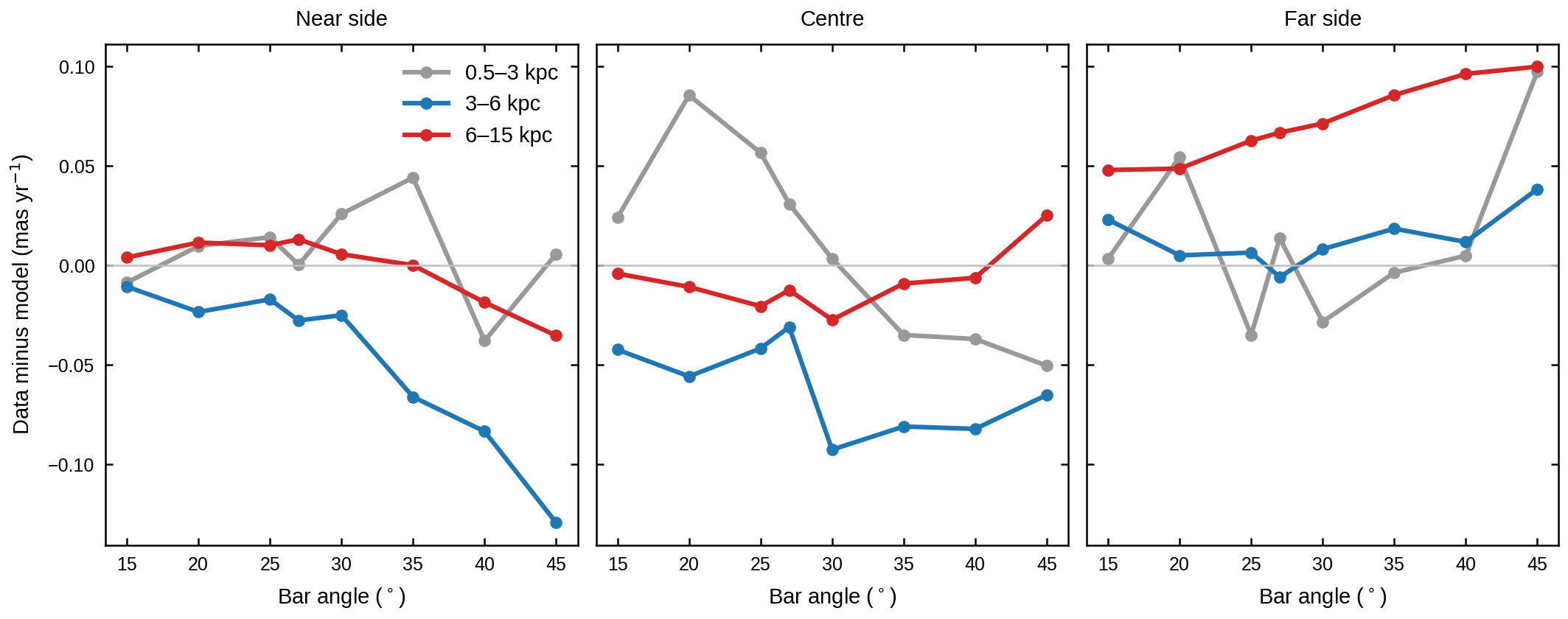}

\noindent\textbf{Extended Data Fig.~7 $|$ Dependence on the bar angle.} Gaia minus the 2.9\,Gyr model, in mas\,yr$^{-1}$, for bar angles from $15^\circ$ to $45^\circ$ on the near side, at the centre and on the far side. Grey, blue and red lines are the 0.5--3, 3--6 and 6--15\,kpc parallax-distance groups. The main comparison uses $27^\circ$.

\clearpage

\noindent\includegraphics[width=\textwidth]{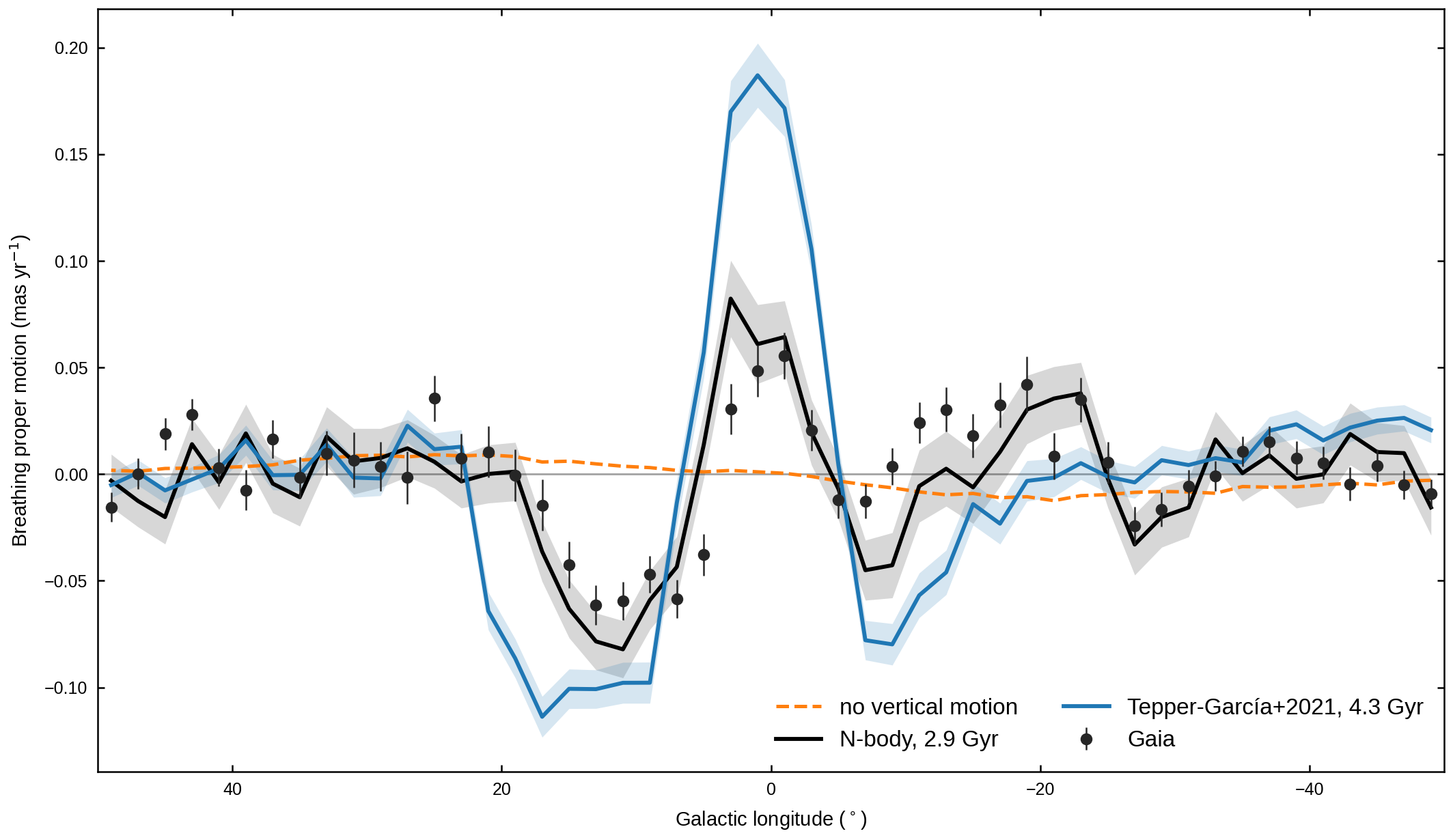}

\noindent\textbf{Extended Data Fig.~8 $|$ Comparison with the public snapshot of the original simulation.} Breathing proper motion against Galactic longitude at parallax distances of 3--15\,kpc. Points are Gaia with $1\sigma$ error bars. The black line is our 2.9\,Gyr model and the blue line is the public snapshot of Tepper-Garc\'ia et al.\cite{TepperGarcia2021} at 4.28\,Gyr (labelled 4.3\,Gyr); shaded bands are their $1\sigma$ sampling uncertainties. The orange dashed line is our model without vertical motion. Values are in mas\,yr$^{-1}$. Longitude increases to the left.

\end{document}